\input epsf
%%%%%%%%%%%%%%%%%%%%%%%%%%%%%%%%%%%%%%%%%%%%%%%%%%%%%%%%%%%%%%%%%
%                                                               %
%       FONT FAMILIES:                                          %
%                                                               %
%%%%%%%%%%%%%%%%%%%%%%%%%%%%%%%%%%%%%%%%%%%%%%%%%%%%%%%%%%%%%%%%%
%                                                               %
%       Define script letters as rsfs                           %
%               (or redefine as cal)                            %
%                                                               %
%                                                               %
%%%%%%%%%%%%%%%%%%%%%%%%%%%%%%%%%%%%%%%%%%%%%%%%%%%%%%%%%%%%%%%%%
\newfam\scrfam
\batchmode\font\tenscr=rsfs10 \errorstopmode
\ifx\tenscr\nullfont
        \message{rsfs script font not available. Replacing with calligraphic.}
        \def\scr{\cal}

\else   
        \font\sevenscr=rsfs7
        \font\fivescr=rsfs5
        \skewchar\tenscr='177 \skewchar\sevenscr='177 \skewchar\fivescr='177
        \textfont\scrfam=\tenscr \scriptfont\scrfam=\sevenscr
        \scriptscriptfont\scrfam=\fivescr
        \def\scr{\fam\scrfam}
        \def\cal{\scr}
\fi
%%%%%%%%%%%%%%%%%%%%%%%%%%%%%%%%%%%%%%%%%%%%%%%%%%%%%%%%%%%%%%%%%
%                                                               %
%       Blackboard bold (or redefine as boldface)               %
%                                                               %
%%%%%%%%%%%%%%%%%%%%%%%%%%%%%%%%%%%%%%%%%%%%%%%%%%%%%%%%%%%%%%%%%
\newfam\msbfam
\batchmode\font\twelvemsb=msbm10 scaled\magstep1 \errorstopmode
\ifx\twelvemsb\nullfont\def\Bbb{\bf}

	\message{Blackboard bold not available. Replacing with boldface.}
\else   \catcode`\@=11
        \font\tenmsb=msbm10 \font\sevenmsb=msbm7 \font\fivemsb=msbm5
        \textfont\msbfam=\tenmsb
        \scriptfont\msbfam=\sevenmsb \scriptscriptfont\msbfam=\fivemsb
        \def\Bbb{\relax\expandafter\Bbb@}
        \def\Bbb@#1{{\Bbb@@{#1}}}
        \def\Bbb@@#1{\fam\msbfam\relax#1}
        \catcode`\@=\active

\fi
%%%%%%%%%%%%%%%%%%%%%%%%%%%%%%%%%%%%%%%%%%%%%%%%%%%%%%%%%%%%%%%%%
%                                                               %
%       MORE FONTS:                                             %
%                                                               %
%%%%%%%%%%%%%%%%%%%%%%%%%%%%%%%%%%%%%%%%%%%%%%%%%%%%%%%%%%%%%%%%%
        \font\eightrm=cmr8              \def\xrm{\eightrm}
        \font\eightbf=cmbx8             \def\xbf{\eightbf}
        \font\eightit=cmti10 at 8pt     \def\xit{\eightit}
%%%     \font\eightit=cmti8             \def\xit{\eightit}
        \font\eighttt=cmtt8             \def\xtt{\eighttt}
        \font\eightcp=cmcsc8
        \font\eighti=cmmi8              \def\xold{\eighti}
        \font\teni=cmmi10               \def\old{\teni}
	\font\twelvei=cmmi12
        \font\tencp=cmcsc10
        \font\tentt=cmtt10
        
        \font\twelvecp=cmcsc10 scaled\magstep1

	\font\eightsym=cmsy8

	 at10pt	
	\font\twelvehelvbold=phvb at12pt
	 at14pt
	\font\sixteenhelvbold=phvb at16pt

\def\noblackbox{\overfullrule=0pt}
\noblackbox

%%%%%%%%%%%%%%%%%%%%%%%%%%%%%%%%%%%%%%%%%%%%%%%%%%%%%%%%%%%%%%%%%
%                                                               %
%       HEADLINE:                                               %
%                                                               %
%%%%%%%%%%%%%%%%%%%%%%%%%%%%%%%%%%%%%%%%%%%%%%%%%%%%%%%%%%%%%%%%%
\newtoks\headtext
\headline={\ifnum\pageno=1\hfill\else
{\eightcp\the\headtext}
                \dotfill{ }{\old\folio}\fi}
\def\makeheadline{\vbox to 0pt{\vss\noindent\the\headline\break
\hbox to\hsize{\hfill}}
        \vskip2\baselineskip}
%%%%%%%%%%%%%%%%%%%%%%%%%%%%%%%%%%%%%%%%%%%%%%%%%%%%%%%%%%%%%%%%%
%                                                               %
%       FOOTNOTES:                                              %
%                                                               %
%%%%%%%%%%%%%%%%%%%%%%%%%%%%%%%%%%%%%%%%%%%%%%%%%%%%%%%%%%%%%%%%%
\newcount\infootnote
\infootnote=0
\def\foot#1#2{\infootnote=1\footnote{$\,{}^{#1}$}{\vtop{\baselineskip=.75\baselineskip\advance\hsize by -\parindent\noindent{\xrm #2}\vskip.4\baselineskip}}\infootnote=0}
%%%%%%%%%%%%%%%%%%%%%%%%%%%%%%%%%%%%%%%%%%%%%%%%%%%%%%%%%%%%%%%%%
%                                                               %
%       REFERENCES:                                             %
%                                                               %
%%%%%%%%%%%%%%%%%%%%%%%%%%%%%%%%%%%%%%%%%%%%%%%%%%%%%%%%%%%%%%%%%
\newcount\refcount
\refcount=1
\newwrite\refwrite
\def\oldsize{\ifnum\the\infootnote=1\xold\else\old\fi}
\def\ref#1#2{
	\def#1{{{\oldsize\the\refcount}}\ifnum\the\refcount=1\immediate\openout\refwrite=\jobname.refs\fi\immediate\write\refwrite{\item{[{\xold\the\refcount}]} 
	#2\hfill\par\vskip-2pt}\xdef#1{{\oldsize\the\refcount}}\global\advance\refcount by 1}
	}
\def\refout{\catcode`\@=11
        \xrm\immediate\closeout\refwrite
        \vskip2\baselineskip
        {\noindent\twelvecp References}\hfill\vskip\baselineskip
                                                %\vskip.25\baselineskip%%%%
        %\parskip=.875\parskip
        %\baselineskip=.8\baselineskip
        \baselineskip=.75\baselineskip
        \input\jobname.refs
        %\parskip=8\parskip \divide\parskip by 7
        %\baselineskip=1.25\baselineskip
        \baselineskip=4\baselineskip \divide\baselineskip by 3
        \catcode`\@=\active\rm}

\def\hepth#1{\href{http://xxx.lanl.gov/abs/hep-th/#1}{{\xtt hep-th/#1}}}
\def\jhep#1#2#3#4{\href{http://jhep.sissa.it/stdsearch?paper=#2\%28#3\%29#4}{J. High Energy Phys. {\xbf #1#2} ({\xold#3}) {\xold#4}}}
\def\PLB#1#2#3{Phys. Lett. {\xbf B#1} ({\xold#2}) {\xold#3}}
\def\NPB#1#2#3{Nucl. Phys. {\xbf B#1} ({\xold#2}) {\xold#3}}

\def\CMP#1#2#3{Commun. Math. Phys. {\xbf #1} ({\xold#2}) {\xold#3}}
\def\JMP#1#2#3{J. Math. Phys. {\xbf #1} ({\xold#2}) {\xold#3}}

\def\JP#1#2#3{J. Phys. {\xbf #1} ({\xold#2}) {\xold#3}}
%%%%%%%%%%%%%%%%%%%%%%%%%%%%%%%%%%%%%%%%%%%%%%%%%%%%%%%%%%%%%%%%%
%                                                               %
%       SECTION NUMBERING:                                      %
%                                                               %
%%%%%%%%%%%%%%%%%%%%%%%%%%%%%%%%%%%%%%%%%%%%%%%%%%%%%%%%%%%%%%%%%
\newcount\sectioncount
\sectioncount=0
\def\section#1#2{\global\eqcount=0
	\global\subsectioncount=0
        \global\advance\sectioncount by 1
        \vskip2\baselineskip\noindent
        \line{\twelvecp\the\sectioncount. #2\hfill}
	\vskip\baselineskip\noindent
        \xdef#1{{\old\the\sectioncount}}}
\newcount\subsectioncount
\def\subsection#1#2{\global\advance\subsectioncount by 1
	\vskip.8\baselineskip\noindent
	\line{\tencp\the\sectioncount.\the\subsectioncount. #2\hfill}
	\vskip.5\baselineskip\noindent
	\xdef#1{{\old\the\sectioncount}.{\old\the\subsectioncount}}}
\newcount\appendixcount
\appendixcount=0
\def\appendix#1{\global\eqcount=0
        \global\advance\appendixcount by 1
        \vskip2\baselineskip\noindent
        \ifnum\the\appendixcount=1
        \hbox{\twelvecp Appendix A: #1\hfill}\vskip\baselineskip\noindent\fi
    \ifnum\the\appendixcount=2
        \hbox{\twelvecp Appendix B: #1\hfill}\vskip\baselineskip\noindent\fi
    \ifnum\the\appendixcount=3
        \hbox{\twelvecp Appendix C: #1\hfill}\vskip\baselineskip\noindent\fi}
\def\acknowledgements{\vskip2\baselineskip\noindent
        \underbar{\it Acknowledgements:}\ }
%%%%%%%%%%%%%%%%%%%%%%%%%%%%%%%%%%%%%%%%%%%%%%%%%%%%%%%%%%%%%%%%%
%                                                               %
%       EQUATION NUMBERING                                      %
%                                                               %
%%%%%%%%%%%%%%%%%%%%%%%%%%%%%%%%%%%%%%%%%%%%%%%%%%%%%%%%%%%%%%%%%
\newcount\eqcount
\eqcount=0
\def\Eqn#1{\global\advance\eqcount by 1
        \xdef#1{{\old\the\sectioncount}.{\old\the\eqcount}}
        \ifnum\the\appendixcount=0
                \eqno({\oldstyle\the\sectioncount}.{\oldstyle\the\eqcount})\fi
        \ifnum\the\appendixcount=1
                \eqno({\oldstyle A}.{\oldstyle\the\eqcount})\fi
        \ifnum\the\appendixcount=2
                \eqno({\oldstyle B}.{\oldstyle\the\eqcount})\fi
        \ifnum\the\appendixcount=3
                \eqno({\oldstyle C}.{\oldstyle\the\eqcount})\fi}
\def\eqn{\global\advance\eqcount by 1
        \ifnum\the\appendixcount=0
                \eqno({\oldstyle\the\sectioncount}.{\oldstyle\the\eqcount})\fi
        \ifnum\the\appendixcount=1
                \eqno({\oldstyle A}.{\oldstyle\the\eqcount})\fi
        \ifnum\the\appendixcount=2
                \eqno({\oldstyle B}.{\oldstyle\the\eqcount})\fi
        \ifnum\the\appendixcount=3
                \eqno({\oldstyle C}.{\oldstyle\the\eqcount})\fi}
\def\multi{\global\advance\eqcount by 1}
\def\multieq#1#2{\xdef#1{{\old\the\eqcount#2}}
        \eqno{({\oldstyle\the\eqcount#2})}}
%%%%%%%%%%%%%%%%%%%%%%%%%%%%%%%%%%%%%%%%%%%%%%%%%%%%%%%%%%%%%%%%%
%                                                               %
%       Hyperrefs:                                        	%
%                                                               %
%%%%%%%%%%%%%%%%%%%%%%%%%%%%%%%%%%%%%%%%%%%%%%%%%%%%%%%%%%%%%%%%%
\newtoks\url
\def\Href#1#2{\catcode`\#=12\url={#1}\catcode`\#=\active#2}
\def\href#1#2{{#2}}

%%%%%%%%%%%%%%%%%%%%%%%%%%%%%%%%%%%%%%%%%%%%%%%%%%%%%%%%%%%%%%%%%
%                                                               %
%       FORMAT:                                                 %
%                                                               %
%%%%%%%%%%%%%%%%%%%%%%%%%%%%%%%%%%%%%%%%%%%%%%%%%%%%%%%%%%%%%%%%%
\parskip=3.5pt plus .3pt minus .3pt
\baselineskip=14pt plus .1pt minus .05pt
\lineskip=.5pt plus .05pt minus .05pt
\lineskiplimit=.5pt
\abovedisplayskip=18pt plus 4pt minus 2pt
\belowdisplayskip=\abovedisplayskip
\hsize=14cm
\vsize=19cm
\hoffset=1.5cm
\voffset=1.8cm
\frenchspacing
\footline={}
%%%%%%%%%%%%%%%%%%%%%%%%%%%%%%%%%%%%%%%%%%%%%%%%%%%%%%%%%%%%%%%%%
%                                                               %
%       VARIOUS DEFINITIONS                                     %
%                                                               %
%%%%%%%%%%%%%%%%%%%%%%%%%%%%%%%%%%%%%%%%%%%%%%%%%%%%%%%%%%%%%%%%%
\def\ss{\scriptstyle}
\def\sss{\scriptscriptstyle}
\def\*{\partial}
\def\punkt{\,\,.}
\def\komma{\,\,,}

\def\+{\!+\!}
\def\={\!=\!}
\def\small#1{{\hbox{$#1$}}}
\def\half{\small{1\over2}}

\def\ie{{\tenit i.e.}}

\def\id{1\hskip-3.5pt 1}

\def\nl{\hfill\break\indent}
\def\nlni{\hfill\break}
%        {\lower2.5pt\hbox{\eightrm #1}\/\raise2.5pt\hbox{\eightrm #2}}}
%\def\ihalf{{\lower2.5pt\hbox{\eightmath i}\/\raise2.7pt\hbox{\eightrm 2}}}
%\def\ifrac#1{{\lower2.5pt\hbox{\eightmath i}\/\raise2.5pt\hbox{\eightrm#1}}}
%\def\Ifrac#1#2{
%{\lower2.5pt\hbox{{\eightrm #1}{\eightmath i}}\/\raise2.5pt\hbox{\eightrm#1}}}

\def\ie{{\tenit i.e.}}

\def\id{1\hskip-3.5pt 1}

\def\nl{\hfill\break\indent}
\def\nlni{\hfill\break}

%%%%%%%%%%%%%%%%%%%%%%%%%%%%%%%%%%%%%%%%%%%%%%%%%%%%%%%%%%%%%%%%%%%%%%%%%%%%%%
%
%     definitions for this paper
%
%%%%%%%%%%%%%%%%%%%%%%%%%%%%%%%%%%%%%%%%%%%%%%%%%%%%%%%%%%%%%%%%%%%%%%%%%%%%%%

\def\L{\Lambda}
\def\G{\Gamma}

\def\R{{\Bbb R}}
\def\C{{\Bbb C}}
\def\H{{\Bbb H}}
\def\O{{\Bbb O}}
\def\K{{\Bbb K}}
\def\Z{{\Bbb Z}}

\def\M{{\cal M}}

\def\II{\hbox{I\hskip-0.6pt I}}

\def\arrover#1{\vtop{\baselineskip=0pt\lineskip=0pt
      \ialign{##\cr$\longrightarrow$ \cr
                \hfill ${\ss #1}$\hfill\cr}}}
		
\def\Re{\hbox{Re}}

\def\first{1${}^{\hbox{\xit st}}$}
\def\second{2${}^{\hbox{\xit nd}}$}

%%%%%%%%%%%%%%%%%%%%%%%%%%%%%%%%%%%%%%%%%%%%%%%%%%%%%%%%%%%%
%
%       THE PAPER
%
%
%%%%%%%%%%%%%%%%%%%%%%%%%%%%%%%%%%%%%%%%%%%%%%%%%%%%%%%%%%%%

\headtext={Martin Cederwall: ``Geometric Construction of 
AdS Twistors''}

\line{
\epsfysize=1.7cm
\epsffile{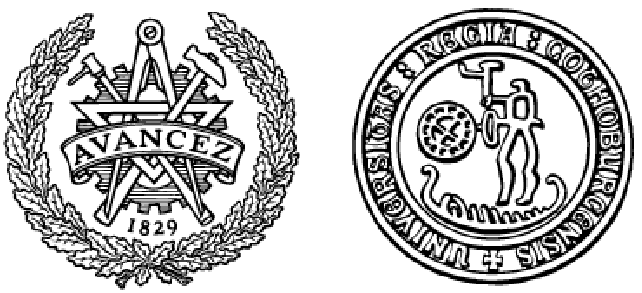}
\hfill}
\vskip-1.7cm
\line{\hfill G\"oteborg ITP preprint}
\line{\hfill\tt hep-th/0002216}
\line{\hfill February, {\old2000}}
\line{\hrulefill}

\vfill

\centerline{\sixteenhelvbold Geometric Construction of AdS Twistors}

\vskip1.6cm

\centerline{\twelvehelvbold Martin Cederwall}

\vskip.8cm

\centerline{\it Institute for Theoretical Physics}
\centerline{\it G\"oteborg University and Chalmers University of Technology }
\centerline{\it S-412 96 G\"oteborg, Sweden}

\vskip1.6cm

{\narrower\noindent 
%Time-like geodesics in AdS${}_{\nu+3}$, $\nu$=1,2,4, are constructed
%geometrically from spinors of the AdS group Spin(2$,\nu$+2), 
\underbar{Abstract:}
Time-like geodesics in AdS${}_{4}$, AdS${}_{5}$ and AdS${}_{7}$ are constructed
geometrically and independently of choice of AdS coordinates
from division algebra spinors of the corresponding AdS groups, 
explaining and generalising the construction by Claus 
{\it et al.} of AdS${}_5$ twistors. 
\smallskip}
\vfill

\line{\hrulefill}
\catcode`\@=11
\line{\tentt martin.cederwall@fy.chalmers.se\hfill}
\catcode`\@=\active
\line{\tentt http://fy.chalmers.se/\~{}tfemc\hfill}

\eject

\ref\Kalloshone{P. Claus, M. G\"unaydin, R. Kallosh, J. Rahmfeld and Y. Zunger,
{\xit ``Supertwistors as quarks of SU(2,2{\eightsym\char'152}4)''},
\nl\jhep{99}{05}{1999}{019} [\hepth{9905112}].}

\ref\Kalloshtwo{P. Claus, R. Kallosh and J. Rahmfeld,
{\xit ``BRST quantization of a particle in AdS${}_5$''},
\hepth{9906195}.}

\ref\Sudbery{A. Sudbery, 
{\xit ``Division algebras, (pseudo)orthogonal groups and spinors''}, 
\JP{A17}{1984}{939}.}

\ref\DivTwistors{I. Bengtsson and M. Cederwall,
{\xit ``Particles, twistors and the division algebras'',}
\nl\NPB{302}{1988}{81}.}

\ref\Maldacena{J. Maldacena,
{\xit ``The large N limit of superconformal field theories and supergravity''},
\nl Int. J. Theor. Phys. {\xbf38} (1999) 1113 [\hepth{9711200}];
\nlni S.S. Gubser, I.R. Klebanov and A.M. Polyakov,
\nl{\xit ``Gauge theory correlators from non-critical string theory''},
\nl\PLB{428}{1998}{105} [\hepth{9802109}];
\nlni E. Witten, {\xit ``Anti de Sitter space and holography''},
\nl Adv. Theor. Math. Phys. {\xbf2} (1998) 253 [\hepth{9802150}].}

\ref\KalloshKilling{R. Kallosh and A.A. Tseytlin,
{\xit ``Simplifying superstring action on AdS${}_{\sss5}$ x S${}^{\sss5}$''},
\nl\jhep{98}{10}{1998}{016} [\hepth{9808088}].}

\ref\Sevensphere{M. Cederwall and C.R. Preitschopf, 
{\xit ``S${}^{\sss7}$ and $\widehat{\hbox{S}^{\sss7}}$''},
\nl\CMP{167}{1995}{373} [\hepth{9309030}].}

\ref\PureSpinors{M. Cederwall, {\xit ``Bi-division algebra spinors in
D=10 Minkowski geometry''}, 
\nl G\"oteborg Preprint {\xold91}-{\xold26} ({\xold1991}).}

\ref\Dimaetal{I. Bandos, J. Lukierski, C. Preitschopf and D. Sorokin,
\nl{\xit ``OSp supergroup manifolds, superparticles and supertwistors''},
\hepth{9907113}.}

\ref\MCten{M. Cederwall, 
{\xit ``Octonionic particles and the S${}^{\sss7}$ symmetry''},
\JMP{33}{1992}{388}. }

\ref\Zunger{Y. Zunger, {\xit ``Twistors and Actions on Coset 
Manifolds''}, \hepth{0001072}}

\ref\MaldacenaOoguri{J. Maldacena and H. Ooguri,
{\xit ``Strings in AdS(3) and SL(2,R) WZW model. 1.''}, \hepth{0001053}.}

\ref\Berkovits{N. Berkovits, 
{\xit ``Super-Poincar\'e covariant quantization of the superstring''}, 
\hepth{0001035}.}

\ref\BerkovitsTwistor{N. Berkovits, 
{\xit ``A supertwistor description of the massless superparticle in
ten-dimensional superspace''}, 
\nl\NPB{350}{1991}{193}.}

\section\Intro{Introduction}In connention with the AdS/CFT correspondence 
[\Maldacena],
there has been a renewed interest in anti-de Sitter (AdS) and sphere
(S) geometries,
supergravities on AdS$\times$S backgrounds, 
and perturbative string theory
on these spaces. In contrast to  flat Minkowski background, where 
perturbative quantisation of strings and NSR superstrings is 
straight-forward (and where supersymmetric quantisation of 
10-dimensional superstrings recently was achieved [\Berkovits]), 
leading to free two-dimensional (super-)conformal field theories,
the situation in AdS space looks extremely difficult, in spite of
the simple structure of the manifold. The 3-dimensional case, where
the bosonic string has been successfully quantised 
[\MaldacenaOoguri], is special in that
AdS${}_{3}$ is a group manifold.

Speculations have occurred whether a twistor formulation might provide
a more natural framework for string quantisation on AdS$\times$S
spaces [\Kalloshtwo]. There is not much substantial evidence for this yet, but
the idea should be tested. In flat space, twistors are associated
with massless particles, since they are spinors under the conformal group.
In AdS space, the types of twistors sofar considered 
[\Kalloshone,\Kalloshtwo,\Dimaetal,\Zunger] transform only under
the AdS isometry group. Therefore, no particular mass is favoured, and
one main argument against the applicability of twistor theory to strings,
namely the observation that string theory is not conformally symmetric,
is absent. Twistor variables may (at least group-theoretically) generate
an entire massive string spectrum.

In spite of some progress, at least to the extent that a superstring action,
with clever choice of fermionic coordinates related to Killing spinors 
[\KalloshKilling],
now looks simpler than some time ago, quantisation is not yet feasible.
The twistor program should be carefully reexamined, and one of the
first things to do is to ask what, in general, an AdS twistor is, and
what kind of equations it satisfies. This is the subject of the present paper.
In the process, a simple geometric definition of the twistor 
variables, independent of choice of coordinates on AdS space and
manifestly covariant under the isometry group, will be obtained. 

The sphere parts of the relevant geometries will be left out of the
present discussion, which consequently means that we pay no attention 
whatsoever
to physically relevant R-symmetries, nor to Kaluza--Klein spectra. 
Neither will supersymmetry
be considered. We regard this paper as setting the framework for 
future investigations including these aspects.

\section\Geodesics{Geometric construction}We start out by demanding that,
on-shell, twistors describe the motion of a massive particle on 
AdS. There is no need to distinguish between AdS and its universal
covering space in this context, since the massive geodesics of the 
covering space are the (infinite) universal coverings of the AdS 
geodesics.

Let us regard AdS${}_{d+1}$ with radius $R$ as the hyperboloid 
$\eta_{MN}X^MX^N=-R^2$
in a flat space with signature $(2,d)$.
The basic observation on which the construction will be based is the
fact that a geodesic on AdS is the intersection between the hyperboloid
and a plane, more specifically a completely time-like plane, through the
origin $X=0$.
The stability group of a time-like plane is SO(2)$\times$SO($d$),
so the space ${\cal M}$ of massive geodesics on AdS${}_{d+1}$ is 
the homogeneous space
$$
\M={\hbox{SO}(2,d)\over\hbox{SO}(2)\times\hbox{SO}(d)}\punkt
\Eqn\GeodesicsSpace
$$
A plane is the type of geometrical object that may naturally be 
parametrised as a spinor (\ie, twistor) bilinear. We will now go through this 
parametrisation, and take it as the operative definition of AdS twistors.

\vskip\baselineskip
\epsfxsize=7.5cm
\centerline{\epsffile{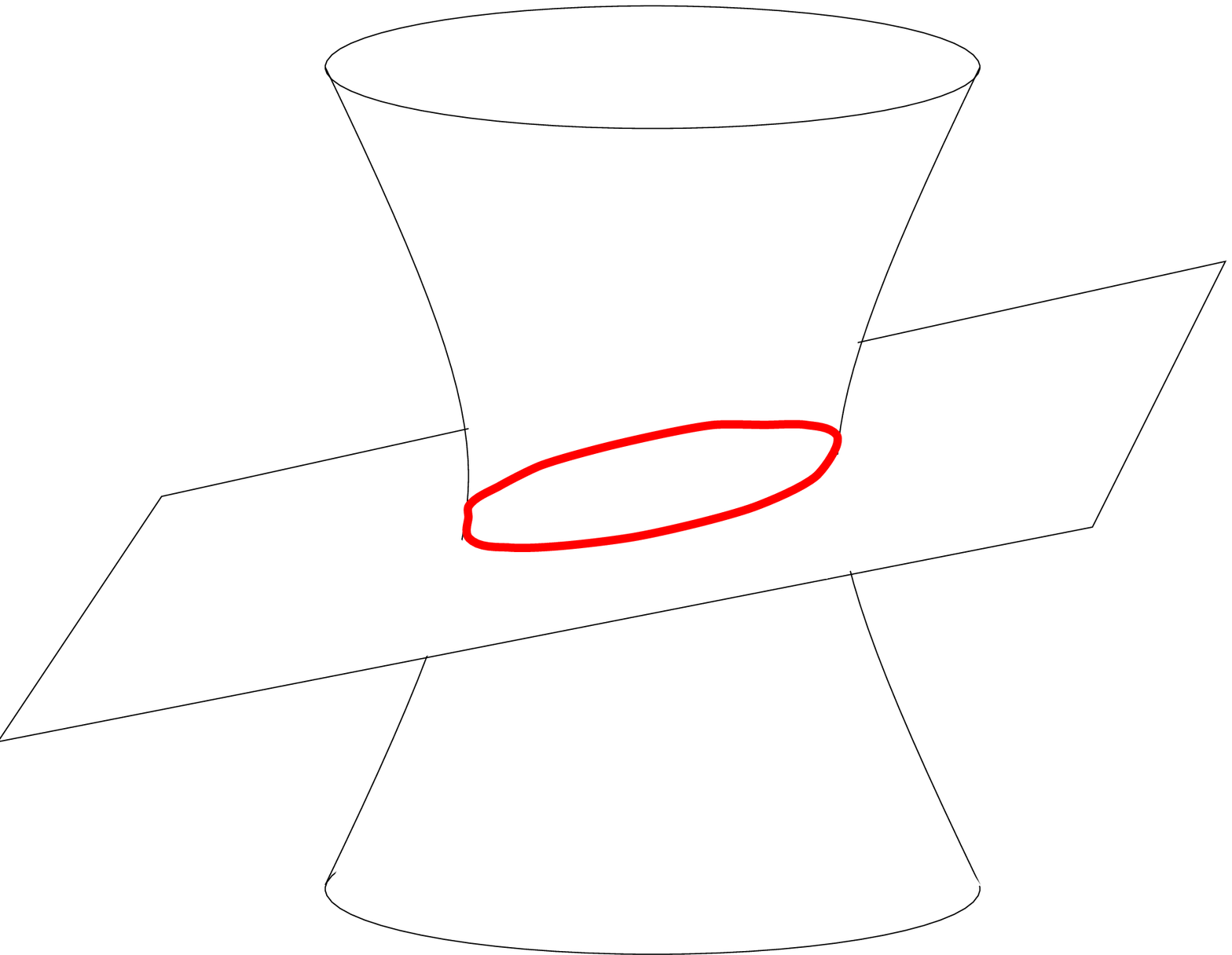}}
\centerline{\it Fig. 1. A geodesic as an intersection with a plane.}
\vskip\baselineskip

When a plane is to be parametrised as a spinor bilinear, the object
formed from a spinor $\L$ that may be used is 
$$
\Pi^{MN}=\half\bar\L\G^{MN}\L\punkt\Eqn\TwistorTransform
$$
This already gives some group-theoretical information, namely that the
symmetric product of two spinor representations of SO$(2,d)$ must contain the
adjoint. All the cases treated here are of this type. In general one
may of course consider other situations, where some 
anti-symmetric/anti-hermitean matrix
is introduced in eq. (\TwistorTransform), but this will at present 
not be needed.

Eq. (\TwistorTransform) will be taken as the basic relation between 
the spinorial and space-time kinematical variables,
the "twistor transform". It is interpreted so that the tensor $\Pi$ defines
the plane in question by spanning directions in the plane. This puts
restrictions on the spinor---in general a tensor formed as in 
eq. (\TwistorTransform)
will not define a single plane. We will find that typically more than
one spinor is needed, so that eq. (\TwistorTransform)
must be seen as a sum over some internal orthogonal index carries by $\L$
(eventually, this index should be a spinor index 
related to the proper R-symmetry group
in an AdS$\times$S setting). A set of spinors fulfilling the adequate
constraints to define a plane through eq. (\TwistorTransform) will
be referred to as {\it simple}. Simple spinors are in a certain sense,
fulfilling homogeneous quadratic constraints, 
quite analogous to pure spinors [\PureSpinors].

We will consider a number of dimensionalities that we find more interesting
than a general setting, partly for their mathematical simplicity and
partly for their physical relevance. The cases AdS${}_4$, AdS${}_5$ and 
AdS${}_7$ are special in that they occur in the supergravity solutions
corresponding to the three non-dilatonic branes: the membrane in
$D=11$, the self-dual 3-brane in type \II B in $D=10$ and the 5-brane
in $D=11$. The corresponding AdS groups SO$(2,3)$, SO$(2,4)$ and SO$(2,6)$
are identical to the conformal groups of 3-, 4- and 6-dimensional 
Minkowski space, and naturally related to the division algebras
$\R$, $\C$ and $\H$, the real, complex and 
quaternionic numbers\foot*{We are not aware of any reason for this 
intriguing
coincidence of physical and mathematical structures---the mathematical       
sequence is natural, but the three M-theory vacua do not share a 
corresponding common origin.}.
We denote the division algebras by $\K_\nu$, $\nu=1,2,4$, and the 
AdS groups (or more precisely the spin groups, their double covers)
are written as Sp$(4;\K_\nu)$ in the sense defined by Sudbery [\Sudbery].
A spinor under the AdS group belongs to the fundamental representation
of Sp$(4;\K_\nu)$, \ie, it is a 4-component column with entries in 
$\K_\nu$.
It transforms in addition under the usual $N=1$ R-symmetry group 
$\Z_2$, U(1) or SU(2) for $\nu=1,2$ or 4 respectively, by right
muliplication by elements of $\K_\nu$ with unit norm. At the algebra level, the
R-symmetry is generated by $A_1(\K_\nu)$, imaginary elements of $\K_\nu$,
which for higher $N$ generalises to $A_N(\K_\nu)$, anti-hermitean 
$N\times N$ matrices. Octonions will not be considered in this
paper---it is straight-forward to check that non-associativity ruins
the corresponding construction for AdS${}_{11}$ (to the initiated reader
we may state this more precisely: the Moufang identities that allows for
the use of $S^7$ as R-symmetry "group" [\Sevensphere] 
for Minkowski (2-component) spinors
do not suffice here---what is demanded is analogous to the existence
of $\O P^3$). 

Let us now investigate the possibility that the twistor transform
(\TwistorTransform), when certain constraints are "imposed", describes
the space (\GeodesicsSpace) of massive geodesics on AdS${}_{\nu+3}$.
To get a better perspective, we would like to compare the present 
construction to what happens for division algebra twistors in Minkowski space
[\DivTwistors,\MCten]. The geometric construction, which in that case consists
of parametrising a light-like vector as a spinor bilinear, then only
involves momenta, and not the coordinates: 
$p^\mu=\bar\lambda\gamma^\mu\lambda$. The momentum is invariant under 
R-symmetry, so this is simply divided out from spinor space to obtain
the configuration space of momenta. This modding out is the Hopf map
$S^{2\nu-1}\arrover{S^{\nu-1}}S^\nu$.
In the present case, the twistor transform involves the entire phase 
space\foot*{Such an interpretation may be given also in Minkowski 
space, if $\ss\Pi$ in eq. (\TwistorTransform) is taken to contain the
generators of conformal transformations. All the interesting geometry
lies in the parametrisation of the momentum, however, since the
space of geodesics is obtained by a direct product with a Cauchy surface.}, 
and the spinor $\L$ will be self-conjugate. 
The symplectic form on twistor phase space is essentially unique, given
the AdS symmetry---it must equal the antisymmetric (or anti-hermitean)
spinor metric\foot\dagger{This equation of course refers to brackets
between {\xit real} components, so in the complex and quaternionic cases
the spinor index is not identical to that of the Appendix, although the
same letters are used.}, $\{\L^A,\L^B\}=\varepsilon^{AB}$.
Therefore, we do not need to derive it from the Poisson
brackets of coordinates and momenta for some specific choice of
AdS coordinates.
The space $\M$ of
geodesics will not be constructed simply as the quotient of a spinor 
space by a symmetry group, but as such a quotient of a constraint 
hypersurface in spinor (this is true whether of not there are
second class constraints---if there is, the dimension of the gauge 
orbit will be smaller than the codimension of the constraint surface.
We will soon see that second class constraints generically {\it are} 
present, with the case of AdS${}_5$ as an exception).

\section\RealSection{The Twistor Transform for Ads${}_{4}$}We must 
determine how many spinors are needed in order to form a 
time-like plane, \ie, how to obtain a simple spinor. 
A straight-forward and practical, but not very 
elegant, way of doing this is by using the AdS group to choose a specific frame
in which the antisymmetric tensor $\Pi$ defining the plane takes 
a particularly simple form.
We let it be $\Pi^{00'}\neq0$ and other components zero, where 0 and $0'$ 
denote the two time directions. We refer to the Appendix for further 
notation. For simplicity, we will do the 4-dimensional (real) case
explicitly, and then comment on the details of AdS${}_5$ and AdS${}_7$.
Using the $\G$-matrices of the appendix, a number of equations 
are obtained that describes the vanishing of all components of $\Pi^{MN}$
except $MN=00'$. It is straight-forward to verify that the minimum
number $N$ of spinors needed is $N=4$. The four spinors are 
conveniently collected in a quaternion, so that
$$
\L=\left[\matrix{\lambda_{1}\cr\lambda_{2}\cr\lambda_{3}\cr\lambda_{4}\cr}
\right]\in\R^4\otimes\H=\H^4\punkt\Eqn\LambdaMatrix
$$ 
The R-symmetry for such spinors is 
SU(2)${}_L\times$SU(2)${}_R$, acting by left and right 
quaternionic multiplication, which obviously commutes with left
multiplication by real 4$\times$4 matrices.
In order for $\bar\L\G^{MN}\L$ to span the $00'$ plane, the algebraic
constraints on the $\lambda_{i}$'s is that they are all orthogonal as
quaternions (SO(4) vectors) and of equal length:
$$
\eqalign{
&\Re(\bar\lambda_{a}\lambda_{b})=0\komma\quad a\neq b\cr
&|\lambda_a|=|\lambda_b|\punkt\cr}
\Eqn\Orthogonality
$$
This is satisfied iff {\it either}
$$
\eqalign{
T_{(L)i}&\equiv\half\Re(\bar\Lambda e_i\Lambda)=0\cr
\hbox{\it or}\quad T_{(R)i'}&\equiv-\half\Re(\bar\Lambda\Lambda 
e_{i'})=0\komma\cr}
\eqn
$$
where $T_{(L,R)}$ are the generators of the R-symmetry SU(2)'s.
This formulation of the constraints is covariant, and must therefore
hold in any frame, not just the one chosen for the calculation.
The two alternatives are mutually exclusive (a spinor fulfilling both
has to vanish), and correspond to the
$\lambda_{a}$'s spanning an orthogonal frame of either orientation.

There is now a problem: only part of the R-symmetry generators (half)
act as algebraic constraints on the spinors, but also the rest need
to be eliminated in order to remove the redundancy of the simple spinor
description of the plane. Let us arbitrarily choose $T_{(L)}=0$. We then
also need a set of algebraic constraints generating SU(2)${}_{R}$,
but without setting $T_{(R)}=0$. The only possibility is $T_{(R)}=v$, 
where $v$ is some non-zero imaginary quaternion.
Specifying $v$ implies that SU(2)${}_{R}$ is broken to U(1).
The unbroken U(1) constraint is \first\ class and
the remaining two constraints are \second\ class.
The physical significance of $v$ is examined in the following section.

We may now check the construction by counting the physical degrees of
freedom. The spinor $\Lambda$ had from the outset 16 real components.
With two \second\ class constraints and four \first\ class ones
(generating SU(2)$\times$U(1)), we
arrive at $16-2-2\times4=6$ degrees of freedom.
This should be checked against the dimensionality of the space
of massive geodesics
$\M=\hbox{SO}(2,3)/(\hbox{SO}(2)\times\hbox{SO}(3))$, which is 
$10-1-3=6$.

It is of course a bit disappointing that \second\ class constraints 
are present. Although they may be eliminated using the Dirac 
procedure without breaking AdS covariance, the resulting Dirac brackets
are $\Lambda$-dependent and do not invite to quantisation. 

\section\GeneralCase{{\twelvei D} = 5 and 7}It is a straight-forward 
exercise to continue the analysis to AdS${}_{5}$ and AdS${}_{7}$.
It is already known [\Kalloshone] that the complex (AdS${}_{5}$)
construction works with two spinors. In the quaternionic case,
four spinors are again needed. This non-uniform pattern is somewhat
surprising---nothing corresponding happens for Minkowski space twistors
(where also the octonionic case works [\BerkovitsTwistor,\MCten]).
Of course, the AdS${}_{5}$ construction also works with four spinors,
but this is a redundant description. We will comment on this below.

A pair of complex spinors has the R-symmetry group 
$A_2(\C)\approx\hbox{SU(2)}\times\hbox{U(1)}$. At the same time, 
the condition that $\Pi^{MN}=\half\bar\L\G^{MN}\L$
spans a plane is identical to the vanishing of the generators of the
SU(2) R-symmetry subgroup, $T_i=0$. The U(1) generator is not allowed
to vanish, but is taken to fulfill an inhomogeneous constraint
$T=v$. Since $T$ generates an abelian subgroup, the set of 
four constraints is nevertheless \first\ class, unlike the real case,
where an inhomogeneous constraint had to be imposed on a non-abelian
part of the R-symmetry generators. The dimensionalities clearly match:
$\hbox{dim}\M=
\hbox{dim}\bigl(\hbox{SO}(2,4)/(\hbox{SO}(2)\times\hbox{SO}(4))\bigr)
=8$, while the number of physical twistor degrees of freedom is
$16-2\times4=8$.

If one choses to use four spinors in the parametrisation of AdS${}_{5}$
geodesics, the R-symmetry is $A_4(\C)\approx\hbox{SU(4)}\times\hbox{U(1)}$.
The entire set of SU(4) generators can not vanish. The R-symmetry
is then broken to SU(2)$\times$SU(2)$\times$U(1)$\times$U(1), with the
rest being \second\ class constraints. Without going into explicit 
detail, the two-spinor case can be recovered by using the
eight \second\ class constraints and simultaneously gauge fixing one
of the SU(2)$\times$U(1) factors, thereby eliminating two of the spinors.

For the quaternionic twistors of AdS${}_{7}$, four spinors are again needed.
The R-symmetry is $A_4(\H)\approx\hbox{Sp(8)}$. Again, only part of 
the generators may fulfill homogeneous constraints, which means that 
there will be a mixture of \first\ and \second\ class constraints.
Here, the condition for $\Pi^{MN}=\half\bar\L\G^{MN}\L$ to span a plane 
implies that Sp(8) has to be broken to SU(4)$\times$U(1), 
of which the U(1) constraint, as usual, is inhomogeneous.
Counting the twistor degrees of freedom from the constraint structure 
gives $64-2\times16-20=12$, matching the dimension of
$\M=\hbox{SO}(2,6)/(\hbox{SO}(2)\times\hbox{SO}(6))$.

What is the meaning of $v$, the length of the non-vanishing 
R-symmetry generator? 
The twistor transform (\TwistorTransform), that described algebraically 
the plane in the flat embedding space, also contains the generators
of the AdS${}_{d+1}$ group Spin(2,$d$). 
The actual geodesic described by the twistors is independent of
the scaling of $\L$, \ie, of $v$.
In this way the choice of $v$ gives the only piece of information about 
particle motion not contained in the geodesic, namely the mass.
The exact relation is easily derived in the frame used above,
where $|\Pi^{00'}|=2|v|$. $\half\Pi^{00'}$ is the generator of translations in
the time variable $\varphi$, the angle about the ``equator'' of AdS, so the
metrically normalised translation generator is 
${1\over2R}\Pi^{00'}$.
We therefore obtain the relation 
$$
m={|v|\over R}\Eqn\VandM
$$
(the numerical factor of course depending on the normalisation of
the corresponding Cartan subalgebra element; the factor in eq. (\VandM)
relates to the explicit formul\ae\ in section \RealSection).

\section\Discussion{Discussion}We have given an explicit and 
geometric division algebra twistor transform 
for the spaces of massive geodesics on
the three AdS spaces corresponding to non-dilatonic branes.

It is noteworthy that the previously 
known construction of twistors for geodesic motion on AdS${}_{5}$
has a more elegant structure than those for AdS${}_{4}$ and AdS${}_{7}$.
The latter two are probably not very useful, since they constain
second class constraints that make the twistor descriptions, 
independently of any actual dynamics, if not intractable, so at least
quite non-linear. They will probably not present any advantages over
the coset description.
It is difficult to refrain from speculating that the particular
properties of AdS${}_5$ twistors might have
something to do with AdS${}_{5}$ arising from string theory in 
$D=10$, while the other two spaces derive from eleven-dimensional 
supergravity/M-theory, although the mathematical reason is simply that
the R-symmetry group in the complex case, and only then, contains a
U(1) factor.

It is of course straight-forward to write down \first-order actions that
reproduce the particle dynamics:
$$
S\sim\int d\tau\Bigl[(\bar\L\dot\L)+\varrho\cdot\bigl(\bar\L\L(T-v)\bigr)\Bigr]
\komma\eqn
$$
where $\varrho$ are Lagrange multipliers.

Apart from the possible application to string quantisation, the 
procedure of which is not obvious, there are two ways the present
formalism has to be extended. The supersymmetric case must be 
addressed, which can probably be done along the same lines as for
supertwistors in Minkowski space, and one will have to consider
geodesic motion on the accompanying sphere manifolds. Hopefully,
the present formalism will extend naturally to a situation where the
R-symmetry acts as isometries of the spheres, and where the twistors,
suitably constrained, simultaneously parametrise planes in the 
embedding spaces of both anti-de Sitter space and the sphere.
The constraint structure will certainly be affected when the R-symmetry
group (or part of it) no longer is gauged, and it is not clear that
the structure found in this paper, with a mixing of \first\ and \second\
class constraints, will persist.
Whether or not the methods introduced here can be useful for
the quantum-mechanical treatment of string theory on AdS spaces
remains to be investigated. The geometric approach can hopefully
help to gain insight in this question.

\acknowledgements
The author would like to thank Renata Kallosh for discussions that 
stimulated this investigation, and also Gary Gibbons and Bengt EW 
Nilsson for clarifying comments.

\vfill\eject

\refout

\vfill\eject

\appendix{Spinors and gamma matrices}Here, we set the conventions
for the gamma matrices used in the twistor transform 
(\TwistorTransform). They can be given in a unified notation for the
three cases. We denote by $e_I$, $I=1,\ldots,\nu$, the standard 
orthonormal basis for the division algebra $\K_\nu$.
For the flat embedding space with signature $(2,\nu+2)$ we use 
light-cone coordinates $M=(\oplus,\ominus,\mu)=(\oplus,\ominus,+,-,I)$ 
and scalar product
$V\cdot W=-V^\oplus W^\ominus-V^\ominus W^\oplus-V^+W^--V^-W^++V^IW^I$.

A spinor under the AdS group belongs to the fundamental representation
of Sp$(4;\K_\nu)$, \ie, it is a 4-component column with entries in 
$\K_\nu$. Using a dotted/undotted notation for spinors, and in addition primed
and unprimed spinor indices (since there generically are two 
chiralities), the gamma matrices (or, strictly speaking, sigma 
matrices) acting on one chirality (the unprimed one that is chosen
for the twistors) are
$$
\eqalign{
&\G^{\oplus A'}{}_B=
\left[\matrix{\sqrt2\,\id^\alpha{}_\beta&0\cr0&0\cr}\right]
\komma\quad
\G^{\ominus A'}{}_B=
\left[\matrix{0&0\cr0&\sqrt2\,\id_{\dot\alpha}{}^{\dot\beta}\cr}\right]
\komma\quad   \cr
&\G^{\mu A'}{}_B=
\left[\matrix{0&\tilde\gamma^{\mu\alpha\dot\beta}\cr
\gamma^{\mu}{}_{\dot\alpha\beta}&0\cr}\right]
\komma   \cr}\eqn
$$
and on the other one
$$
\eqalign{
&\tilde\G^{\oplus A}{}_{B'}=
\left[\matrix{0&0\cr0&-\sqrt2\,\id_{\dot\alpha}{}^{\dot\beta}\cr}\right]
\komma\quad
\tilde\G^{\ominus A}{}_{B'}=
\left[\matrix{-\sqrt2\,\id^\alpha{}_\beta&0\cr0&0\cr}\right]
\komma\quad  \cr
&\tilde\G^{\mu A}{}_{B'}=
\left[\matrix{0&\tilde\gamma^{\mu\alpha\dot\beta}\cr
\gamma^{\mu}{}_{\dot\alpha\beta}&0\cr}\right]
\komma  \cr}\eqn
$$
where $\gamma^\mu$, $\tilde\gamma^\mu$ are 
SL(2;$\K_\nu$) $\approx$ Spin(1,$\,\nu+1$) gamma 
matrices:
$$
\eqalign{
&\gamma^+{}_{\dot\alpha\beta}=\left[\matrix{\sqrt2&0\cr0&0\cr}\right]
\komma\quad
\gamma^-{}_{\dot\alpha\beta}=\left[\matrix{0&0\cr0&\sqrt2\cr}\right]
\komma\quad
\gamma^I{}_{\dot\alpha\beta}=\left[\matrix{0&\bar e_I\cr e_I&0\cr}\right]
\cr
&\tilde\gamma^{+\alpha\dot\beta}=\left[\matrix{0&0\cr0&-\sqrt2\cr}\right]
\komma\quad
\tilde\gamma^{-\alpha\dot\beta}=\left[\matrix{-\sqrt2&0\cr0&0\cr}\right]
\komma\quad
\tilde\gamma^{I\alpha\dot\beta}=\left[\matrix{0&\bar e_I\cr e_I&0\cr}\right]
\punkt\cr}\eqn
$$
The matrices $\Gamma^{MN}$ used in the construction of the plane 
defining the geodesics are constructed as
$$
\Gamma^{MNA}{}_B=
\half\bigl(\tilde\Gamma^M\Gamma^N-\tilde\Gamma^N\Gamma^M\bigr)^A{}_B
\komma\eqn
$$
and the twistor bilinear is given by 
$$
\Pi^{MN}=\half\bar\L\G^{MN}\L\equiv\half\bar\L_A\Gamma^{MNA}{}_B\L^B
\equiv\half\L^{\dagger\dot A}\varepsilon_{\dot AB}\Gamma^{MNB}{}_C\L^C
\komma\eqn
$$
where the anti-hermitean ``spinor metric''
$$
\varepsilon_{\dot AB}=
\left[\matrix{0&\id_{\dot\alpha}{}^{\dot\beta}\cr
-\id^\alpha{}_\beta&0\cr}\right]
\eqn
$$ 
is used to lower the spinor index, and
where $\dagger$ implies division algebra conjugation.
In the real case, dots are of course superfluous, and there is only 
one chirality. The above formul\ae\ are still correct, and primed and
unprimed indices are then related via
$$
E^{A'}{}_B=\left[\matrix{0&\varepsilon^{\alpha\beta}\cr
\varepsilon_{\alpha\beta}&0\cr}\right]
\punkt\eqn
$$
\end